\documentclass[11pt]{article}
\usepackage{amsmath, amsthm, amssymb, amsfonts}
\usepackage{algorithm}
\usepackage[noend]{algpseudocode}
\usepackage{graphicx}
\usepackage{authblk}
\usepackage{sidecap}
\usepackage{fullpage}
\usepackage{color}

\newcommand{\cA}{\ensuremath{\mathcal{A}}}

\newcommand{\cB}{\ensuremath{\mathcal{B}}}

\newcommand{\cI}{\ensuremath{\mathcal{I}}}

\newcommand{\cK}{\ensuremath{\mathcal{K}}}
\newcommand{\cL}{\ensuremath{\mathcal{L}}}

\newcommand{\cO}{\ensuremath{\mathcal{O}}}

\newcommand{\cP}{\ensuremath{\mathcal{P}}}

\newcommand{\cT}{\ensuremath{\mathcal{T}}}

\newcommand{\bbC}{\ensuremath{\mathbb{C}}}

\newcommand{\bbG}{\ensuremath{\mathbb{G}}}

\newcommand{\bbP}{\ensuremath{\mathbb{P}}}

\newcommand{\bbR}{\ensuremath{\mathbb{R}}}

\newcommand{\bbS}{\ensuremath{\mathbb{S}}}
\newcommand{\bbT}{\ensuremath{\mathbb{T}}}

\DeclareMathOperator{\Bd}{\partial}

\DeclareMathOperator{\Div}{Div}

\DeclareMathOperator{\Ostar}{Ost}

\DeclareMathOperator{\bary}{Bary}

\DeclareMathOperator{\inte}{Int}

\DeclareMathOperator{\names}{names}%mph changed ids->names
\DeclareMathOperator{\skel}{skel}

\DeclareMathOperator{\views}{views}
\DeclareMathOperator{\view}{view}
\newcommand{\var}[1]{\lstinline+#1+}

\theoremstyle{remark}
\newtheorem{theorem}{Theorem}[section]

\newtheorem{definition}[theorem]{Definition}

\newtheorem{fact}[theorem]{Fact}

\newtheorem{lemma}[theorem]{Lemma}

\newtheorem{property}[theorem]{Property}

\newcommand{\set}[1]{\{#1\}}

\newcommand{\send}{\textbf{send}}
\newcommand{\recv}{\textbf{recv}}

\newcommand{\validated}{\mathrm{Val}}

\newcommand{\ctI}{\tilde{\cI}}
\newcommand{\ctO}{\tilde{\cO}}
\newcommand{\tDelta}{\tilde{\Delta}}
\newcommand{\csI}{\cI^*}
\newcommand{\csO}{\cO^*}
\newcommand{\sDelta}{\Delta^*}

\algblockdefx[upon]{Upon}{EndUpon}%
	[1]{\textbf{upon} #1 \textbf{do}}%
	{}
\graphicspath{{.}}

\title{\bf Distributed Computability in Byzantine Asynchronous Systems}
\author[a]{Hammurabi Mendes}
\author[b]{Christine Tasson}
\author[a]{Maurice Herlihy\footnote{Supported by NSF 000830491.}}

\affil[a]{
      Computer Science Dept.\\
      Brown University\\
      Providence, RI, USA\\
      \texttt{\{hmendes,mph\}@cs.brown.edu}
}
\affil[b]{
      Univ. Paris Diderot, Sorbonne Paris Cit\'e\\
      PPS, UMR 7126, CNRS, F-75205\\
      Paris, France\\
      \texttt{Christine.Tasson@pps.univ-paris-diderot.fr}
}
\date{}

\begin{document}

\maketitle

``\copyright 2014 Hammurabi Mendes, Christine Tasson, Maurice Herlihy. This is the author's version of the work. It is posted here for your personal use. Not for redistribution. The definitive Version of Record was published in STOC'14. (will be available at \texttt{http://dx.doi.org/10.1145/2591796.2591853}).''

\begin{abstract}
In this work, we extend the topology-based approach for characterizing computability
in asynchronous crash-failure distributed systems
to asynchronous Byzantine systems.
We give the first theorem with necessary and sufficient conditions
to solve arbitrary tasks in asynchronous Byzantine systems
where an adversary chooses faulty processes.
In our adversary formulation,
outputs of non-faulty processes are constrained in terms of inputs
of non-faulty processes only.
For colorless tasks,
an important subclass of distributed problems,
the general result reduces to an elegant model
that effectively captures the relation between
the number of processes,
the number of failures,
as well as the topological structure of the task's simplicial complexes.
\end{abstract}

\section{Introduction}
\label{Sec-Introduction}

A \emph{task} is a distributed coordination problem involving multiple computing \emph{processes}.
Each process starts with a private input,
taken from a finite set,
communicates with other processes,
and eventually decides on a private output,
also taken from a finite set.
One of the central questions in distributed computing is characterizing which
tasks can be solved in which models of computation.
Those models specify
synchrony, communication, and failure characteristics/guarantees.
A \emph{protocol} is a distributed algorithm that solves a task
given a model of computation.

In this work,
we consider asynchronous systems,
where processes have different relative speeds,
and communication is subject to unbound, yet finite delays.
We are interested in situations where a subset of \emph{faulty} processes can exist.
Two failure models are often discussed in the literature.
In well-studied \emph{crash-failure} models~\cite{DSBookLynch},
faulty processes simply halt without warning, at possibly different times.
In the more severe \emph{Byzantine-failure} models~\cite{Lamport1982},
faulty processes can behave arbitrarily, even maliciously.
Here,
we address the problem of characterizing which tasks are solvable
in asynchronous Byzantine systems.
Even though necessary and sufficient conditions for computability
in crash-failure systems have long been known~\cite{HerlihyShavit1999},
this work provides the first general computability characterization
for asynchronous Byzantine systems.

\subsection{Our Contributions}
\label{Sec-Contributions}

Tools adapted from combinatorial topology have been successful in
characterizing task solvability in synchronous and asynchronous
\emph{crash-failure} systems,
as in~\cite{HerlihyShavit1999}.
This paper extends the approach
to tasks in asynchronous Byzantine systems~\cite{Lamport1982,DSBookLynch}.
The results presented here suggest
that the language of combinatorial topology (a generalization of the language of graphs)
is a convenient and effective way to formalize a range of distinct distributed computing models.
We present some background on distributed computing in Sec.~\ref{Sec-OperationalModel},
and outline its topology-based modeling in Sec.~\ref{Sec-TopologicalModel}.

Our principal contribution, presented in Sec~\ref{Sec-EquivalenceTheorem},
is to give the \emph{first theorem}
with necessary and sufficient conditions to solve arbitrary tasks
in asynchronous Byzantine systems.
In our approach,
a Byzantine-failure task is defined in terms of a pair of combinatorial structures called
\emph{simplicial complexes}~\cite{Munkres84,Kozlov07},
and a map modeling task semantics.
We assume an adversary that may deem a subset of processes as faulty,
and require that the output of non-faulty processes is permitted
in light of the input of non-faulty processes,
according to the task's formal specification.
Our theorem says that,
in asynchronous systems,
a Byzantine-failure task is solvable
if and only if
a \emph{dual} crash-failure task,
also expressed in terms of simplicial complexes,
is solvable.
Given that solvability conditions have long been known for crash failures
(see~\cite{HerlihyShavit1999}),
our equivalence theorem, presented in Sec.~\ref{Sec-EquivalenceTheorem},
provides for the first time solvability conditions for Byzantine failures
in asynchronous systems.

Furthermore,
the above characterization reduces to a particularly elegant, compact form for
\emph{colorless tasks}~\cite{BorowskyGLR01,HerlihyR12},
an important subclass of tasks that encompasses well-studied problems such as
consensus~\cite{FLP}, $k$-set agreement~\cite{ksetagreement},
and approximate agreement~\cite{approx-consensus,MendesHerlihy13}.
For those tasks,
we provide additional results in Sec.~\ref{Sec-ColorlessTasks} capturing the relation between
the number of processes,
the number of failures,
and the topological structure of the task's simplicial complexes.

\subsection{Related Work}
\label{Sec-RelatedWork}

The Byzantine failure model was initially introduced by Lamport,
Shostak, and Pease in~\cite{Lamport1982}.
Most of the literature in this area has focused on the synchronous model (survey in~\cite{FischerConsensusSurvey}),
not on the (more demanding) asynchronous model considered here.
The reliable broadcast protocol is adapted from
Bracha~\cite{Bracha} and from Srikanth and Toueg~\cite{SriTouRB}.
Malkhi \emph{et al.}~\cite{MalkhiMRT2003} propose several computational models
in which processes communicate via shared objects (instead of messages),
and display Byzantine failures.
De Prisco \emph{et al.}~\cite{dePriscoMR2001} consider the $k$-set agreement problem
in a variety of asynchronous settings.
Their notion of the validity condition for the $k$-set agreement problem, however,
is weaker than ours.
Neiger~\cite{Neiger93}
discusses a stronger validity condition similar to the constraints used here.

\section{Operational Model}
\label{Sec-OperationalModel}

We have $n+1$ processes\footnote{
Choosing $n+1$ processes rather than $n$ simplifies the topological notation,
but also slightly complicates the computing notation.
Choosing $n$ processes has the opposite trade-off.
We choose $n+1$ for compatibility with prior work.
}
$P_0, \ldots, P_n$,
that exchange messages via pairwise distinct channels.
These channels form a complete graph,
and are \emph{reliable} and \emph{FIFO}:
all transmitted messages are eventually delivered, in the order they were sent.
Communication is \emph{asynchronous}:
the delivery of any message happens after a finite,
yet unbounded delay.
Given that the channels are pairwise distinct,
the sender of any message is reliably identified
(\emph{authenticated channels} in the literature~\cite{RSPBook}).
The processes are asynchronous as well,
with unbound relative speed.

Up to $t$ processes might be \emph{faulty}.
We discuss two classical models for process failures.
In the \emph{crash-failure} model~\cite{DSBookLynch},
processes execute the protocol as prescribed,
yet the faulty processes become permanently silent at any point in the execution
(i.e., halt and never send additional messages).
In the \emph{Byzantine-failure} model \cite{Lamport1982},
faulty processes display arbitrary, even malicious behavior,
which includes collusion to prevent the protocol to terminate correctly.
In a crash-failure system,
the faulty processes are precisely the ones that halt during the execution.
In our Byzantine-failure system, however,
the faulty processes are a set of no more than $t$ processes
chosen by an \emph{adversary}.
Byzantine processes may execute the protocol correctly or incorrectly,
at the discretion of the adversary.
Regarding notation,
the set of all processes is denoted by $\bbP$,
partitioned in non-faulty processes $\bbG \subseteq \bbP$
and faulty processes $\bar{\bbG} = \bbP \setminus \bbG$.

We model processes as state machines.
Each process $P_i$ has an internal state, or \emph{view}, written $\view(P_i)$.
Initially, $\view(P_i)$ is the process' starting input.
In this work,
we are interested in task solvability,
but not in communication complexity
(i.e., number and size of messages).
Hence,
we assume that processes follow a \emph{full-information} protocol~\cite{HerlihyRT09}.
Each process repeatedly:
(1) receives the state information from other processes;
(2) concatenates that information to its own internal state;
(3) sends its internal state to all other processes.
After completing some number of iterations,
each process applies a decision function $\delta$ to its current state in order to decide.

As a first measure in this work,
we use higher-level communication abstractions
on top of the bare message-passing model.
These new abstractions are described below.

\subsection{Reliable Broadcast}
\label{Sec-ReliableBroadcast}

\emph{Reliable broadcast} is a well-known technique that forces Byzantine processes to
communicate consistently with other processes~\cite{Bracha,SriTouRB,DSBook,RSPBook}.
The communication is organized in asynchronous rounds,
where a round may involve several message exchanges.
Messages have the form $(P,r,c)$,
where $P$ is the sending process,
$r$ is the current round, and $c$ is the actual content.
Messages not conforming to this structure can safely be discarded.
The technique, which works as long as $n + 1 > 3t$,
guarantees the following:
\begin{description}
\item[Non-Faulty Integrity:]
If a non-faulty $P$ never reliably broadcasts $(P,r,c)$,
then no non-faulty process ever reliably receives $(P,r,c)$.
\item[Non-Faulty Liveness:]
If a non-faulty $P$ reliably broadcasts $(P,r,c)$,
then all non-faulty processes will reliably receive $(P,r,c)$ eventually.
\item[Global Uniqueness:]
If two non-faulty processes $Q$ and $R$ reliably receive,
respectively, $(P,r,c)$ and $(P,r,c')$,
then the messages are equal ($c = c'$),
even if the sender $P$ is Byzantine.
\item[Global Liveness:]
For two non-faulty processes $Q$ and $R$,
if $Q$ reliably receives $(P,r,c)$,
then $R$ will reliably receive $(P,r,c)$ eventually,
even if the sender $P$ is Byzantine.
\end{description}

We overview the algorithms
in Appendix~\ref{App-ReliableBroadcast}.
For details, please refer to~\cite{DSBook,RSPBook,Bracha}.

In this text,
$P.\mathtt{RBSend}(M)$ denotes the reliable broadcast of message $M$ by process $P$,
and $P.\mathtt{RBRecv}(M)$ the reliable receipt of $M$ by $P$.
Unless otherwise noted,
all messages exchanged in asynchronous Byzantine systems use reliable broadcast.

%\subsection{Adversaries and Schedules}
%\label{Sec-AdversariesSchedules}

%We assume an adversarial scheduler that controls:
%\begin{enumerate}
%\item
%the delivery of any message previously sent
%the instant when the message becomes available to be accepted);
%\item
%in a Byzantine system, which are the Byzantine processes;
%\item
%in a Byzantine system, the arbitrary behavior of the Byzantine processes.
%\end{enumerate}

%An \emph{action} is either:
%(a) steps 1 to 3 above for some non-faulty process;
%(b) the delivery of any message sent by some non-faulty process;
%(c) the sending and delivery of an arbitrary message from some Byzantine process.
%A \emph{schedule} is a total order of \emph{actions}.

\section{Topological Model}
\label{Sec-TopologicalModel}

We now overview some important notions from combinatorial topology,
and describe how they model concurrent computation.
For details, please refer to Munkres~\cite{Munkres84} or Kozlov \cite{Kozlov07}.

\subsection{Combinatorial Tools}
\label{Sec-CombinatorialTools}

A \emph{simplicial complex} $\cK$ consists of a finite set $V$
along with a collection of subsets of $V$ closed under containment.
An element of $V$ is called a \emph{vertex} of $\cK$.
Each set in $\cK$ is called a \emph{simplex},
usually denoted by lower-case Greek letters: $\sigma, \tau$, etc.
A subset of a simplex is called a \emph{face}.
The \emph{dimension} $\dim(\sigma)$ of a simplex $\sigma$ is $|\sigma|-1$.
We use ``$k$-simplex'' as shorthand for ``$k$-dimensional simplex'',
also in ``$k$-face''.
The dimension $\dim(\cK)$ of a complex is the maximal dimension of its simplices.
The set of vertices of $\cK$ is denoted by $V(\cK)$.
The set of simplices of $\cK$
having dimension at most $\ell$ is a~subcomplex of $\cK$,
which is called \emph{$\ell$-skeleton} of $\cK$,
denoted by $\skel^\ell(\cK)$.

Let $\cK$ and $\cL$ be complexes.
A \emph{vertex map} $f$ carries vertices of $\cK$ to vertices of $\cL$.
If $f$ additionally carries simplices of $\cK$ to simplices of $\cL$,
it is called a \emph{simplicial} map.
A~\emph{carrier map} $\Phi$ from $\cK$ to $\cL$ takes each
simplex $\sigma\in\cK$ to a subcomplex $\Phi(\sigma) \subseteq \cL$,
such that for all $\sigma,\tau \in \cK$,
we have $\Phi(\sigma \cap \tau) \subseteq \Phi(\sigma) \cap \Phi(\tau)$.
A simplicial map $\phi: \cK \to \cL$ is \emph{carried by the carrier map}
$\Phi: \cK \to 2^{\cL}$ if, for every simplex $\sigma \in \cK$,
we have $\phi(\sigma) \subseteq \Phi(\sigma)$.

Although we defined simplices and complexes in a purely combinatorial way,
they can also be interpreted geometrically.
An $n$-simplex can be identified with the convex hull of $(n+1)$
affinely-independent points in the Euclidean space of appropriate dimension.
This geometric realization can be extended to complexes.
The point-set that underlies such \emph{geometric complex} $\cK$ is called
the \emph{polyhedron} of $\cK$, denoted by $|\cK|$.

We can define simplicial and carrier maps between geometrical complexes.
Given a simplicial map $\phi: \cK \to \cL$ (resp. carrier map $\Phi: \cK \to 2^{\cL}$),
the polyhedrons of every simplex in $\cK$ and $\cL$
induce a continuous simplicial map $\phi_c: |\cK| \to |\cL|$ (resp. continuous carrier map $\Phi_c: |\cK| \to |2^{\cL}|$).
We say that $\phi$ (resp. $\phi_c$) is carried by $\Phi$
if, for every simplex $\sigma \in \cK$,
we have $|\phi(\sigma)| \subseteq |\Phi(\sigma)|$ (resp. $\phi_c(|\sigma|) \subseteq \Phi_c(|\sigma|)$).

\subsection{Tasks in the Crash Failure Model}
\label{Sec-TasksCrashFailureModel}

We now present the formalization of crash-failure tasks
as in~\cite{HerlihyR10Shellable}.
In this work,
the input value (resp. output value) of process $P_i$ will always be denoted by $I_i$ (resp. $O_i$).

\begin{definition}
\label{definition-labeled}
A \emph{name-labeled} simplex $\sigma$ is a simplex where:
\begin{enumerate}
	\item for any vertex $v \in \sigma$ we have $v = (P_i,V_i)$ with $P_i \in \bbP$;
	\item if $(P_i,V_i) \in \sigma$ and $(P_j,V_j) \in \sigma$ then $P_i \ne P_j$.
\end{enumerate}
\end{definition}

\begin{definition}
For any name-labeled simplex $\sigma$,
\begin{align*}
\names(\sigma) & = \set{P_i: \exists V \textrm{ such that } (P_i,V) \in \sigma} \mathrm{,} \\
\views(\sigma) & = \set{V_i: \exists P \textrm{ such that } (P,V_i) \in \sigma} \mathrm{.}
\end{align*}
\end{definition}

\begin{definition}
For any simplicial complex $\cK$,
\begin{displaymath}
\names(\cK) = \bigcup_{\sigma \in \cK} \names(\sigma) \textrm{ and } 
\views(\cK) = \bigcup_{\sigma \in \cK} \views(\sigma) \mathrm{.}
\end{displaymath}
\end{definition}

\begin{definition}
\label{definition-canonical}
A \emph{canonical} simplex $\sigma$ is a simplex with $\dim(\sigma) \ge n - t$.
\end{definition}

\begin{definition}
\label{definition-match}
Two simplices $\sigma_1$ and $\sigma_2$ \emph{match} if $$\names(\sigma_1) = \names(\sigma_2) \mathrm{.} $$
\end{definition}

\begin{definition}
\label{definition-namepreserving}
A carrier map $\Phi: \cK \to 2^{\cL}$ is \emph{name-preserving}
if, for any $\sigma \in \cK$, also including vertices of $\cK$,
$\names(\sigma) = \names(\Phi(\sigma))$.
%$\sigma$ matches any $\tau \in \Phi(\sigma)$.
\end{definition}

For any crash-failure task $\cT$,
an \emph{initial configuration} for $\cT$ is
$\sigma_I = \set{(P_i,I_i): I_i \textrm{ is input by } P_i}$,
a canonical name-labeled simplex describing a permitted input to $\cT$.
A \emph{final configuration} for $\cT$ is
$\sigma_O = \set{(P_i,O_i): O_i \textrm{ is output by } P_i}$,
a canonical name-labeled simplex describing a permitted output of $\cT$.
Any initial or final configuration is a canonical simplex
because at least $(n+1)-t$ non-faulty processes
start and finish every computation of $\cT$.

\begin{definition}
\label{definition-crashfailure}
A crash-failure \emph{task specification} is formally a triple $\cT = (\cI,\cO,\Delta)$ such that:
\begin{itemize}
	\item $\cI$ is the \emph{input complex}. A simplex $\sigma \in \cI$ if there~is some $\sigma_I \supseteq \sigma$
that is an initial configuration, with $\dim(\sigma_I) = n$.
	\item $\cO$ is the \emph{output complex}. A simplex $\sigma \in \cO$ if there is some $\sigma_O \supseteq \sigma$
that is a final configuration, with $\dim(\sigma_O) \ge n - t$.
	\item $\Delta: \cI \to 2^{\cO}$ is a name-preserving carrier map.
The simplex $\tau \in \Delta(\sigma)$ if the final configuration $\tau$ is valid given the initial configuration $\sigma$, with $\sigma$ matching $\tau$.
Also, $\Delta(\sigma') = \emptyset$ for any non-canonical simplex $\sigma'$ of~$\cI$.
\end{itemize}
\end{definition}
If an initial configuration $\sigma$ has $\Delta(\sigma) = \emptyset$,
that is, having no associated final configuration,
it effectively precludes any protocol.
Therefore,
tasks are usually defined such that
$\Delta(\sigma) \ne \emptyset$ for any initial configuration $\sigma$.

\subsection{Tasks in the Byzantine Failure Model}
\label{Sec-TasksByzantineFailureModel}

In Byzantine-failure tasks,
we only care about the relation between inputs and outputs of the non-faulty processes.
The task's outputs should be consistent,
despite the participation (sometimes even correct)
of the Byzantine processes,
a property sometimes called \emph{strong validity}, as in~\cite{Neiger93}.
We consider an adversarial model in which
\emph{any set} of up to $t$ processes may be chosen as faulty,
with those processes displaying arbitrary behavior.
Regardless of which processes are faulty,
any final configuration of the \emph{non-faulty processes}
must be permitted in respect to
the initial configuration of the \emph{non-faulty processes}.
For that goal,
our task specification $\cT = (\cI,\cO,\Delta)$
constrains the behavior of non-faulty processes only.

Consider a Byzantine-failure task, say $\cT$.
A \emph{non-faulty initial configuration} for $\cT$ is a canonical name-labeled simplex
$\sigma_{I} = \set{(P_i,I_i) \textrm{ with } P_i \in \bbG}$,
capturing the attribution of inputs to non-faulty processes.
Additionally,
a \emph{non-faulty final configuration} for $\cT$ is a canonical name-labeled simplex
$\sigma_{O} = \set{(P_i,O_i) \textrm{ with } P_i \in \bbG}$
capturing the attribution of outputs to non-faulty processes.

Importantly,
in our adversarial context,
\emph{any set} of no more than $t$ processes may be deemed as faulty.
This adversarial power implies the following property:
\begin{property}
\label{property-adversarypower}
If $\sigma$ is a non-faulty initial configuration:
\begin{enumerate}
	\item There exists a simplex $\sigma^n \supseteq \sigma$ with $\dim(\sigma^n) = n$
where $\sigma^n$ is a non-faulty initial configuration;
	\item For all $\sigma' \subseteq \sigma$ with $\dim(\sigma') \ge n - t$,
we have that $\sigma'$ is a non-faulty initial configuration.
\end{enumerate}
\end{property}

The formal task definition for Byzantine tasks practically mirrors the definition for crash-failure tasks,
although it only constrains task semantics for non-faulty processes:
\begin{definition}
\label{definition-byzantinefailure}
A Byzantine-failure \emph{task specification} is formally a triple $\cT = (\cI,\cO,\Delta)$ such that:
\begin{itemize}
	\item $\cI$ is the \emph{input complex}. A simplex $\sigma \in \cI$ if there~is some $\sigma_I \supseteq \sigma$
that is a non-faulty initial configuration, with $\dim(\sigma_I) = n$.
	\item $\cO$ is the \emph{output complex}. A simplex $\sigma \in \cO$ if there is some $\sigma_O \supseteq \sigma$
that is a non-faulty final configuration, with $\dim(\sigma_O) \ge n - t$.
	\item $\Delta: \cI \to 2^{\cO}$ is a name-preserving carrier map.
The simplex $\tau \in \Delta(\sigma)$ if the non-faulty final configuration $\tau$ is valid given the non-faulty initial configuration $\sigma$, with $\sigma$ matching $\tau$.
Also
$\Delta(\sigma') = \emptyset$ for any non-canonical simplex $\sigma'$ of~$\cI$.
\end{itemize}
\end{definition}

The map $\Delta$ could in principle be
other than a carrier map.
However,
for the sake of studying task computability,
it is enough to assume $\Delta$ as being a carrier map.
Consider the following scenario,
where
$\Delta(\sigma_1 \cap \sigma_2) \not \subseteq \Delta(\sigma_1) \cap \Delta(\sigma_2)$,
for~$\sigma_1,\sigma_2 \in \cI$.
Also consider an asynchronous protocol,
with~$\names(\sigma_1 \cap \sigma_2)$ representing all non-faulty processes,
and~$\names(\sigma_1 \setminus \sigma_2)$ as well as~$\names(\sigma_2 \setminus \sigma_1)$
representing faulty processes.
Now,
the adversary can suitably delay the faulty processes
so that non-faulty processes cannot discern
if the non-faulty initial configuration was~$\sigma_1$, $\sigma_2$, or~$(\sigma_1 \cap \sigma_2)$.
Since we assumed an asynchronous protocol,
a decision must be made,
and it should be inside~$\Delta(\sigma_1 \cap \sigma_2)$ in order to cover all possibilities.

With that in mind,
take a Byzantine task $\cT_1 = (\cI,\cO,\Delta)$,
with $\Delta$ being an arbitrary map,
and another task $\cT_2 = (\cI,\cO,\Delta')$,
identical to $\cT_1$ except that
$\Delta' \subseteq \Delta$ is a carrier map:
$\Delta'(\sigma_1 \cap \sigma_2) \subseteq \Delta'(\sigma_1) \cap \Delta'(\sigma_2)$
for any $\sigma_1,\sigma_2 \in \cI$.
As we have seen,
the \emph{unrestricted} task $\cT_1$
is solvable only if
its \emph{constrained} task $\cT_2$
is solvable.
In addition,
if the constrained task is solvable,
then clearly
the unconstrained task is solvable.
Therefore,
for the sake of studying task solvability,
the map in the Byzantine task definition is simply a carrier map.

\subsection{Pseudospheres}
\label{Sec-Pseudospheres}

Informally,
a \emph{pseudosphere} is a simplicial complex formed by assigning values to processes independently,
such that a process $P_i$ receives only values from a set $S_i$.
In many problems,
input and output complexes are pseudospheres or unions of pseudospheres.

Formally,
a pseudosphere is denoted by $\cK = \Psi(\Sigma,\bbS)$,
where $\bbS \subseteq \bbP$
represents the processes to which we attribute values,
and $\Sigma = \set{\textrm{non-name-labeled } \sigma_i \textrm{ for each } P_i \in \bbS}$
defines such values.
Vertices of $\sigma_i$ are independently assigned to the corresponding process $P_i$.
A name-labeled simplex $\sigma \in \Psi(\Sigma,\bbS)$ if
$\sigma = \set{(P_i,v_i): P_i \in \bbS \textrm{ and } v_i \in \sigma_i \in \Sigma}$.
When $\sigma_i = \sigma_j = \sigma$ for all $i \ne j$,
we call the resulting pseudosphere a \emph{simple pseudosphere},
written as $\Psi(\sigma; \bbS)$.

\section{The Equivalence Theorem}
\label{Sec-EquivalenceTheorem}

In this section,
we will present our main theorem.
We show that, in asynchronous $t$-resilient systems,
a task $\cT_b = (\cI,\cO,\Delta)$ is solvable in the Byzantine failure model
if and only if its \emph{dual} task $\cT_c = (\ctI,\ctO,\tDelta)$ is solvable in the crash failure model,
with $\ctI$, $\ctO$, and $\tDelta$
suitably and respectively defined in terms of $\cI$, $\cO$, and $\Delta$.
We call the task $\cT_b$ the \emph{Byzantine counterpart} of $\cT_c$.

The use of reliable broadcast avoids the problem where
Byzantine processes deliberately send conflicting information to different processes.
However, Byzantine processes
can still introduce a false input and execute the protocol correctly,
yet selectively delaying or omitting certain messages.
Each non-faulty process must choose a correct output
even though a Byzantine process is indistinguishable from
a non-faulty process having an authentic input.
If the input complex is not a simple pseudosphere,
two or more Byzantine processes can introduce incompatible inputs,
and each non-faulty process must decide correctly,
without necessarily detecting which,
if any,
of the incompatible inputs was indeed authentic.
Our dual crash-failure task must be able to capture such issues.

\subsection{Defining the Dual Task}
\label{Sec-DefiningDualTask}

We now formally define the dual task $\cT_c = (\ctI,\ctO,\tDelta)$
in terms of the Byzantine task $\cT_b = (\cI,\cO,\Delta)$.

\begin{definition}
\label{definition-initialconf-byzcounterpart}
An initial configuration of $\cT_c$ is a name-labeled simplex $\sigma_I \in \ctI$
such that it contains a canonical name-labeled simplex $\sigma_G \in \cI$,
with $\views(\sigma_I) \subseteq \views(\cI)$.
\end{definition}

\begin{definition}
\label{definition-finalconf-byzcounterpart}
A final configuration of $\cT_c$ is a name-labeled simplex $\tau_O \in \ctO$
such that it contains a canonical name-labeled simplex $\tau_G \in \cO$,
with $\views(\tau_O) \subseteq \views(\cO)$.
\end{definition}

\begin{definition}
\label{definition-byzantinecounterpart}
Given a Byzantine task $\cT_b = (\cI, \cO, \Delta)$,
its \emph{dual} crash-failure task $\cT_c = (\ctI, \ctO, \tDelta)$ is such that:
\begin{itemize}
	\item $\ctI$ is the \emph{input complex}. A simplex $\sigma \in \ctI$ if there is some $\sigma_I \supseteq \sigma$
that is a possible initial configuration of~$\cT_c$, with $\dim(\sigma_I) = n$.
	\item $\ctO$ is the \emph{output complex}. A simplex $\sigma \in \ctO$ if there is some $\sigma_O \supseteq \sigma$
that is a possible final configuration of~$\cT_c$, with $\dim(\sigma_O) \ge n-t$.
	\item $\tDelta: \ctI \to 2^{\ctO}$ is a name-preserving carrier map, defined below.
\end{itemize}
\end{definition}
Particularly, if $\cI$ (respectively $\cO$) is a simple pseudosphere with dimension $n$,
then $\ctI = \cI$ (respectively $\cO = \ctO$).

\begin{definition}
\label{definition-tdelta}
The map $\tDelta$ is a name-preserving carrier map where:
\begin{multline}
\label{equation-tdelta}
\tDelta(\sigma) = \set{\tau \in \ctO: \forall \textrm{ canonical } \sigma' \subseteq \sigma \textrm{ with } \sigma' \in \cI,\\
\exists \textrm{ matching } \tau' \subseteq \tau \textrm{ with } \tau' \in \cO \textrm{ and } \tau' \in \Delta(\sigma')} \mathrm{,}
\end{multline}
or $\emptyset$ if $\sigma$ is not a canonical simplex in $\ctI$.
Intuitively, the map $\tDelta$ satisfies the original Byzantine specification for any possible choice of
non-faulty processes.
\end{definition}

The map $\tDelta$ is defined as a carrier map,
so $$\tDelta(\sigma_1 \cap \sigma_2) \subseteq \tDelta(\sigma_1) \cap \tDelta(\sigma_2)$$
for any $\sigma_1,\sigma_2 \in \ctI$.
Also,
$\tDelta(\sigma') = \emptyset$ for any non-canonical simplex $\sigma'$ of $\ctI$,
satisfying the final constraint on $\tDelta$ imposed by crash-failure tasks.

%Note that
%if $\Delta(\sigma_1 \cap \sigma_2) = \emptyset$ then $\tDelta(\sigma_1 \cap \sigma_2) = \emptyset$
%for any $\sigma_1,\sigma_2 \subseteq \cI$,
%given the definition of~$\tDelta$.
%If $\sigma_1 \cap \sigma_2$ is a canonical simplex of~$\cI$,
%it is also an initial configuration in the dual task
%(Property~\ref{property-adversarypower} and Definition~\ref{definition-initialconf-byzcounterpart}),
%but the initial configuration has no associated final configuration.
%This has an interesting interpretation.
%Having no associated final configuration for an initial configuration means that
%the dual task is not solvable.
%However,
%the Byzantine task is not solvable either:
%the adversary may suitably delay faulty processes such that
%non-faulty processes cannot tell whether the input started in~$\sigma_1$ or~$\sigma_2$.
%Nevertheless,
%they must be able to decide on this particular scenario
%in order to define a protocol.

\subsection{Solvability Correspondence}
\label{Sec-SolvabilityCorrespondence}

Given an algorithm for $\cT_c$,
we construct an algorithm for $\cT_b$,
showing that,
in asynchronous, $t$-resilient systems,
if we solve the crash-failure $\cT_c = (\ctI, \ctO, \tDelta)$
then we also solve its Byzantine counterpart $\cT_b = (\cI,\cO,\Delta)$

For that purpose,
each non-faulty process $P_i$ will maintain a table $T_i$.
The table has one entry for each combination of process and round:
for process $p \in \bbP$ and round $r \ge 1$,
the contents of the corresponding entry at $T_i$ is denoted as $T_i(p,r)$.
An unfilled entry has contents $\bot$.

In the first round,
non-faulty processes exchange values in $\views(\cI)$.
For any non-faulty processes $P_i \in \bbG$ and $P_j \in \bbP$,
the entry for $T_i(P_j,1)$ contains $(P_j, v)$ only if $(P_j, v) \in V(\cI)$.
Note that
$T_i(P_i, 1) = (P_i, I_i)$,
representing the process' own input (Lines~\ref{algCE:sendinput}-\ref{algCE:sendinputEND}).
For any process set $\bbS$,
$T_i(\bbS, r)$ is the set containing $T_i(p,r)$ iff $p \in \bbS$.

\begin{definition}
\label{definition-starting}
A \emph{starting process set} $\bbS \subset \bbP$ for $P_i \in \bbG$
is a set where the simplex $\sigma = \set{e: e \in T_i(\bbS,1)}$ is an initial configuration of~$\cT_c$.
\end{definition}

In subsequent rounds,
non-faulty processes exchange sets of size at least $(n + 1) - t$.
Such sets satisfy some requirements at the (non-faulty) sender, before being transmitted,
and at the (non-faulty) receiver, before being accepted.
More specifically,
consider $P_i \in \bbG$.
The entry for $T_i(P_i,r)$ is set to $(P_i, V)$
as soon as the predicate $\validated_i(P_i, r, V)$ becomes true.
Then,
the \emph{corresponding message} for $T_i(P_i,r)$, namely $(P_i, r, V)$,
is sent (Lines~\ref{algCE:makesetsMID}-\ref{algCE:makesetsEND}).
Moreover,
if some $P_j \in \bbP$ sends $(P_j, r, V)$,
as soon as the message reaches $P_i$
and the predicate $\validated_i(P_j, r, V)$ becomes true,
the message is accepted by $P_i$.
Then,
the \emph{corresponding entry} for $(P_j, r, V)$, namely $T_i(P_j, r)$,
is set to $(P_j, V)$ (Lines~\ref{algCE:recvsets}-\ref{algCE:recvsetsEND}).
%The validation predicate is formalized below.

\begin{definition}
\label{definition-validated}
The predicate $\validated_i(p, r, V)$ evaluates to true \emph{only if}:
\begin{enumerate}
	\item If $r = 1$, then $(p, V) \in V(\cI)$;
	\item If $r > 1$, then
	\begin{enumerate}
		\item $|V| \ge (n + 1) - t$;
		\item $T_i(p, r - 1) \in V$;
		\item $V \subseteq T_i(\bbS, r - 1)$, for some starting process set $\bbS$ for $P_i$.
	\end{enumerate}
\end{enumerate}
\end{definition}

In Algorithm~\ref{Alg-ConstrainedExecution},
we present the protocol solving $\cT_b$,
as run by non-faulty processes.
The parameter $I_i$ is $P_i$'s input,
and $\cP$ is a crash-failure protocol for $\cT_c$, which is identical across non-faulty processes.
The loop at Line~\ref{algCE:loop} runs until a particular set,
an $R_i$-consistent process set $\bbC_i$, is found.
The definition follows below.

\begin{definition}
\label{definition-consistent}
An $R$-\emph{consistent process set} $\bbS \subseteq \bbP$ for $P_i \in \bbG$ is one where
\begin{enumerate}
		\item $\bbS$ is a starting process set;
		\item $T_i(p,r) \ne \bot$ for all $p \in \bbS$ and $1 \le r \le R$;
		\item $P_i \in \bbS$.
\end{enumerate}
\end{definition}

By definition,
for any $R$-consistent process set $\bbS$,
we have that $\sigma = \set{e: e \in T_i(\bbS,1)}$
is an initial configuration of~$\cT_c$,
with $|\bbS| \ge (n + 1) - t$ and all messages accepted and validated
up to the asynchronous round $R$.
A \emph{decidable} $R$-consistent process set $\bbS$
is one where $\cP$,
if simulated only with entries of $T_i(\bbS, 1 \ldots R) = \set{T_i(\bbS, r): 1 \le r \le R}$,
returns an output.

\begin{algorithm}[htb]
\caption{$P_i.\mathrm{ConstrainedExecution}(I_i, \cP)$}
\label{Alg-ConstrainedExecution}
\begin{algorithmic}[1]
\State By default $T_i(p,r) \leftarrow \bot$ for all $p \in \bbP$ and $r \ge 1$

\State $T_i(P_i,1) \leftarrow (P_i, I_i)$ \label{algCE:sendinput}
\State $\mathtt{RBSend}((P_i, 1, I_i))$ \label{algCE:sendinputEND}

\While{$\not \exists$ decidable $R_i$-consistent process set $\bbC_i$} \label{algCE:loop}
	\Upon{First $V$ with $\validated_i(P_i, r, V)$ and $r > 1$} \label{algCE:makesets}
		\State $T_i(P_i, r) \leftarrow (P_i, V)$ \label{algCE:makesetsMID}
		\State $\mathtt{RBSend}((P_i, r, V))$ \label{algCE:makesetsEND}
	\EndUpon

	\Upon{$\mathtt{RBRecv}((P_j, r, V))$ with $\validated_i(P_j, r, V)$} \label{algCE:recvsets}
		\State $T_i(P_j, r) \leftarrow (P_j, V)$ \label{algCE:recvsetsEND}
	\EndUpon
\EndWhile

\State simulate $\cP$ using only entries of $T_i(\bbC_i, 1 \ldots R)$ \label{algCE:run}
\State \Return own decision value from the above execution
\end{algorithmic}
\end{algorithm}

We also assume that non-faulty processes keep processing messages
as in Lines~\ref{algCE:makesets}~to~\ref{algCE:recvsetsEND}
even after exiting the loop of Line~\ref{algCE:loop}.
This can be seen as a kind of background service interleaved with the steps of the protocol,
similar to~\cite{AttiyaRenaming}\footnote{
To make sure these extra messages are compatible with $\cP$,
as soon as $P_i \in \bbG$ reaches Line~\ref{algCE:run},
its $(R_i + 1)$-th round message contains only entries $T_i(\bbC_i, R_i)$,
``announcing'' the decision.
The execution of~$\cP$ could then ignore $P_i$'s messages accordingly.
}.

\begin{lemma}
\label{lemma-accnonfaultymsgs}
If some $P_i \in \bbG$ fills $T_i(p,r)$ with $(p,V)$,
then any other $P_j \in \bbG$ eventually fills $T_j(p,r)$ with $(p,V)$.
\end{lemma}
\begin{proof}
By induction on $r$.
\textbf{Base:} $r = 1$.
If $P_i$ fills $T_i(p,1)$ with $(p,V)$,
then $\validated_i(p, 1, V)$ is true,
implying that $(p,v) \in V(\cI)$,
by (1)~in Definition~\ref{definition-validated}.
By the liveness properties of the reliable broadcast,
the corresponding message $(p, 1, V)$ eventually reaches any other $P_j \in \bbG$,
and $\validated_j(p, 1, V)$ will be true for identical reason,
filling $T_j(p,1)$ with $(p,V)$.

\textbf{IH:}
Assume that for all $r' < r$,
if some $P_i \in \bbG$ fills $T_i(p,r')$ with $(p,V)$,
then any other $P_j \in \bbG$ eventually fills $T_j(p,r')$ with $(p,V)$.
If $P_i$ fills $T_i(p,r)$ with $(p,V)$,
then $\validated_i(p, r, V)$ is true.
Hence,
$P_i$ filled all entries $T_i(\bbS, r - 1)$,
considering $\bbS$ as the set in (2)-(c) on Definition~\ref{definition-validated}.
By the induction hypothesis,
all those entries are eventually filled in any other $P_j \in \bbG$,
which eventually makes $\validated_j(p, r, V)$ to be true.
By the liveness properties of the reliable broadcast,
the corresponding message $(p, r, V)$ eventually reaches $P_j$,
filling $T_j(p,r)$ with $(p,V)$.
\end{proof}

The entries in the tables on non-faulty processes represent
an asynchronous, $t$-resilient, crash-failure execution schedule for~$\cP$.
Crash failure processes execute the full-information protocol described in Sec.~\ref{Sec-OperationalModel}.
For any asynchronous round $r > 0$,
we interpret $T_i(P_j, r) = (P_j, V)$,
as the scenario where $P_j \in \bbP$ broadcasts $(P_j, r, V)$ and $P_i \in \bbG$ receives $(P_j, r, V)$.
By the previous lemma,
if a message is received by some non-faulty process,
it is eventually received by any other non-faulty process.
Moreover,
by the validation predicate,
$T_i(P_j, r) \ne \bot$ implies in $T_i(P_j, r - 1) \ne \bot$ for all $r > 1$.
An empty entry in $T_i$ represents the inherent inability of a crash-failure processes $P_i$
to discern between a failed process and a sender whose message is delayed.

\begin{lemma}
\label{lemma-exists:consistent}
Any non-faulty process $P_i \in \bbG$ eventually reaches Line~\ref{algCE:run},
and its simulation returns an output.
\end{lemma}
\begin{proof}
We show that
$\bbG$ is bound to be recognized as an $r$-consistent process set at $P_i$
for any $r \ge 1$.
\textbf{Base:} $r = 1$.
Non-faulty processes execute the protocol correctly, so,
by the previous lemma,
$T_i(\bbG, 1)$ is eventually filled.
\textbf{IH:}
Now assume that $T_i(\bbG, r - 1)$ is totally filled.
Fix some $P_i \in \bbG$.
By the previous lemma,
all non-faulty process $P_j \in \bbG$ eventually have $\validated_j(P_j, r, T_j(\bbG, r - 1))$ as true,
sending a message $(P_j, r, V)$ for some $V$,
although not necessarily with $V = T_j(\bbG, r - 1)$.
All those messages are eventually delivered and accepted by $P_i$,
again by the previous lemma,
and $T_i(\bbG, r)$ is eventually filled.
We conclude that $\bbG$ is bound to be recognized as an $r$-consistent process set at $P_i$.

If $P_i$ considers solely the entries of an $r$-consistent process set to simulate an execution of $\cP$,
we actually denote a valid $t$-resilient, asynchronous, crash-failure schedule for $\cP$.
This simulates an initial configuration of $\cT_c$,
under the perspective of $P_i$,
up to the asynchronous round $r$.
Hence,
there exists a concrete $R_i > 0$ such that
$P_i$ reaches Line~\ref{algCE:run} with some \emph{decidable}
$R_i$-consistent process set $\bbC_i$,
although not necessarily $\bbC_i = \bbG$,
with $\cP$ returning an output.
\end{proof}

\begin{lemma}
\label{lemma-CESolvesByzantine}
If an asynchronous, $t$-resilient crash-failure protocol $\cP$ solves $\cT_c = (\ctI,\ctO,\tDelta)$,
then Algorithm~\ref{Alg-ConstrainedExecution} solves its Byzantine counterpart
$\cT_b = (\cI,\cO,\Delta)$.
\end{lemma}
\begin{proof} 
Take some $P_i \in \bbG$, calling $R_i = r$.
We say that $p_1 \in \bbP$ is \emph{seen} by $P_i$ on its execution of~$\cP$ at Line~\ref{algCE:run}
if there is a sequence $p_1 \ldots p_r$ such that
$(p_\ell, V_\ell)$ in $T_i(p_{\ell + 1}, \ell + 1)$ for all $1 \le \ell < r$, and $p_r \in \bbC_i$.
Let
$\sigma_i = \set{T_i(p,1): p \textrm{ is seen by} P_i}$
be the input observed by $P_i$ on its execution of~$\cP$ at Line~\ref{algCE:run}.

A Byzantine process $P_b$ such that $(P_b, v) \in \sigma_i$ for some $P_i \in \bbG$
is said to have \emph{apparent input} $v$.
As all messages are validated through the validation predicate,
we have that $v \in V(\cI)$,
in light of (2)-(1) in Definition~\ref{definition-validated}.

The non-faulty inputs define $\sigma_G \in \cI$,
and the non-faulty plus apparent inputs define
$$ \sigma_A = \bigcup_{P_p \in \bbG} \sigma_i \mathrm{.} $$
The input observed by $P_i$, $\sigma_i$, is an initial configuration of~$\ctI$
including $P_i$,
by Definitions~\ref{definition-consistent} and~\ref{definition-starting}.
Therefore,
the simulation of~$\cP$ at Line~\ref{algCE:run} produces an output in
$\tDelta(\sigma_i) \ne \emptyset$ for all $P_i \in \bbG$.
In addition,
$\sigma_G \subseteq \sigma_A$,
by definition of $\sigma_A$,
and apparent inputs are in $V(\cI)$,
as discussed before.
Then, $\sigma_A$ is an initial configuration of $\cT_c$ as well,
and $\tDelta(\sigma_A) \ne \emptyset$.

By the previous lemma,
recalling our assumption that $\cP$ is a crash-failure protocol for~$\cT_c$,
any non-faulty process $P_i \in \bbG$ will reach Line~\ref{algCE:run},
producing an output $O_i$ such that
\begin{equation}
\label{equation-outputpi}
(P_i,O_i) \in \tau_i \subseteq \tau \mathrm{,}
\end{equation}
with $\tau_i \in \tDelta(\sigma_i)$ and $\tau \in \tDelta(\sigma_A)$.

Since simulations in Line~\ref{algCE:run}
run considering partial views of
a global asynchronous, $t$-resilient, crash-failure schedule for $\cP$,
decisions are consistent among non-faulty processes
(or we contradict the fact that $\cP$ solves $\cT_c$ in asynchronous, $t$-resilient, crash-failure systems).
In other words,
we have that $(P_j,O_j) \in \tau_j \subseteq \tau$, with $\tau_j \in \tDelta(\sigma_j)$
and with the same $\tau$ as in~(\ref{equation-outputpi}), above.
In other words,
\begin{align*}
\tau_G = \set{(P_i,O_i): P_i \in \bbG} \subseteq \tau \in \tDelta(\sigma_A) \mathrm{,}
\end{align*}
with $\tau_G \in \Delta(\sigma_G)$, by definition of $\tDelta$.
As the choice of non-faulty process is totally arbitrary,
the protocol solves $\cT_b$.
\end{proof}

\begin{theorem}
\label{theorem-equivalence}
In asynchronous, $t$-resilient systems,
the task $\cT_b = (\cI,\cO,\Delta)$ is solvable in the Byzantine failure model
if and only if
the task $\cT_c = (\ctI,\ctO,\tDelta)$ is solvable in the crash failure model.
\end{theorem}
\begin{proof}
\textbf{$\cT_b$ implies $\cT_c$.}
Consider an execution of $\cT_b$
where Byzantine processes may only fail by crashing,
having inputs in $\views(\cI)$.
Non-faulty and apparent inputs,
those pertaining to Byzantine processes,
define $\sigma \in \ctI$.

At least one canonical simplex having values in $\cI$ exists by definition of $\ctI$.
Given a protocol for $\cT_b$,
for any canonical simplex $\sigma' \subseteq \sigma$ with $\sigma' \in \cI$,
effectively representing non-faulty processes,
their outputs $\tau'$ are such that $\tau' \in \Delta(\sigma')$,
in order to satisfy any adversarial definition of non-faulty processes.
Of course, $\tau'$ matches $\sigma'$,
and the protocol is actually computing $\tDelta$.
The implication follows because any Byzantine protocol is also a crash protocol.

\textbf{$\cT_c$ implies $\cT_b$.}
Follows from Lemma~\ref{lemma-CESolvesByzantine}.
\end{proof}

In the following section,
we present applications of our Equivalence Theorem
in the context of colorless tasks.
We particularly remark
one important consequence:
for certain colorless tasks,
we will have a lower bound on the number of processes in order to allow these tasks to be solvable.
This number, furthermore,
is expressed in terms of the number of failures
and the task's simplicial complexes.

\section{Colorless Tasks}
\label{Sec-ColorlessTasks}

Colorless tasks~\cite{HerlihyR12,BorowskyGLR01} consist of an important class of problems
where tasks are totally defined in terms of the input and output \emph{sets} of values,
not the particular attribution of values to processes.
In this section,
a specific application of the topological tools yields novel and elegant computability results for
colorless tasks.

A classical example of colorless task is $k$-set agreement~\cite{ksetagreement}.
Say that processes start with input values from a finite set $V$.
In crash-failure systems,
informally speaking,
a protocol solves $k$-set agreement if outputs satisfy:
(1) Agreement -- no more than $k$ different outputs exist; and
(2)	Validity -- any output was proposed in the input.
In Byzantine systems,
a natural variation,
called \emph{strict} $k$-set agreement,
requires that processes decide on values proposed by non-faulty processes \emph{only}
(as in~\cite{Neiger93}, for consensus).

Task specifications can be simplified and specialized for colorless tasks.
We are then able to express computability results in a more elegant language.

\subsection{Model}
\label{Sec-Model}

For colorless tasks,
input and output simplices represent \emph{sets} of input and output values
permitted in the initial/final configurations.
Such sets are closed under inclusion, that is,
if $S$ represents a valid initial (resp. final) input (resp. output) set,
so is any $S' \subseteq S$.
The admissible \emph{sets} of output values
depend solely on
the \emph{set} of input values
taken by processes.
Importantly,
for any set of values,
any particular attribution of those values to processes is valid.

In light of these properties,
we give simpler task specification for colorless tasks.
Initially, we consider crash failures,
following the model in~\cite{HerlihyR10}.
The Byzantine specification will essentially mirror the approach
of Sec.~\ref{Sec-TasksByzantineFailureModel}.

A \emph{colorless task} is a triple $(\csI,\csO,\sDelta)$,
where $\csI$ is the \emph{colorless input complex},
$\csO$ is the \emph{colorless output complex},
and $\sDelta: \csI \to 2^{\csO}$ is the \emph{colorless carrier map}.
Each vertex in $\csI$ (resp. $\csO$) is a possible input (resp. output) value,
and each simplex is a possible \emph{initial input (resp. output) set}.
Given an initial input set,
$\sDelta$ specifies which final output sets are legal.
Colorless tasks can of course be expressed in the general model $(\cI,\cO,\Delta)$,
as seen in~\cite{HerlihyR10}:
\begin{align}
\label{eqn-eqvcoloredcolorless1}
\sigma^* \in \csI \textrm{ (resp. $\csO$) } & \Leftrightarrow
                                           \Psi(\sigma^*;\bbP) \subseteq \cI \textrm{ (resp. $\cO$)} \\
\label{eqn-eqvcoloredcolorless2}
\tau \in \Delta(\sigma) & \Leftrightarrow \views(\tau) \in \sDelta(\views(\sigma))
\end{align}

For instance, for the $(t+1)$-set agreement,
the map $\sDelta$ is the skeleton operator $\skel^t$
as output values must be chosen among at most $t+1$ different values.

With Byzantine tasks,
$\csI$ (resp. $\csO$) refers to \emph{non-faulty} input (resp. output) sets only.
Since the adversary may choose any set with up to $t$ processes as Byzantine
(including the empty set),
the following relation remains valid:
$\sigma^* \in \csI$ (resp. $\csO$) if and only if $\Psi(\sigma^*;\bbP) \subseteq \cI$ (resp. $\cO$).
The map $\sDelta$ is defined as in Equation~\ref{eqn-eqvcoloredcolorless2},
which now conditions non-faulty output sets to non-faulty input sets only.

Before proceeding with computability results specific to colorless tasks,
we define some additional topological tools.

\subsection{Simplicial Approximations}
\label{Sec-SimplicialApproximations}

\begin{figure}[!htb]
\centering{\includegraphics[width=0.6\textwidth]{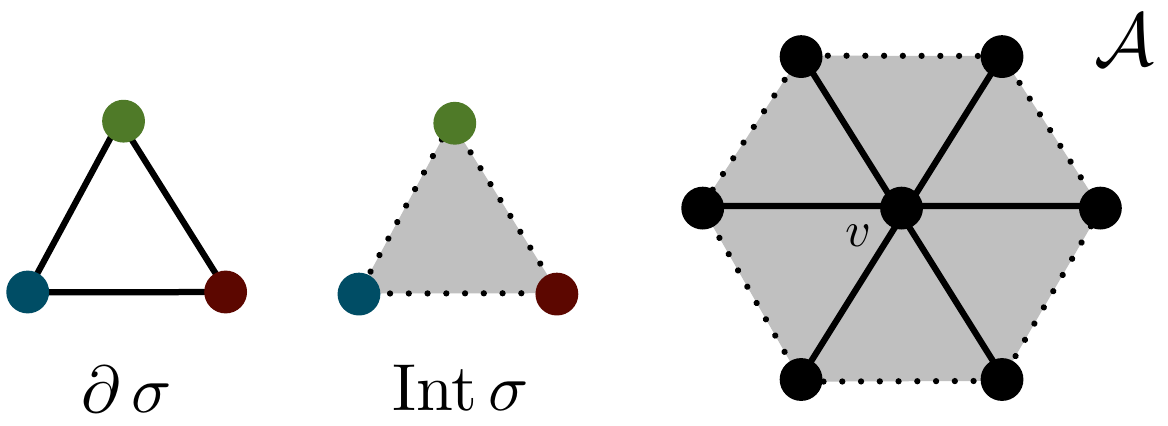}}
\caption{The interior and boundary of a $2$-simplex $\sigma$, and the open star of a $0$-simplex $\set{v} \subseteq \cA$.}
\label{Fig-StandardConstructions}
\end{figure}

Recall that simplicial complexes can be viewed under
a combinatorial or a geometrical perspective.
Combinatorially speaking,
for any simplex $\sigma$, the \emph{boundary} of $\sigma$, denoted $\Bd{\sigma}$,
is the simplicial complex of $(\dim(\sigma)-1)$-faces of $\sigma$.
Geometrically speaking,
the \emph{interior} of $\sigma$ is formally defined as
$\inte{\sigma} = |\sigma| \setminus |\Bd{\sigma}|$.
The \emph{open star} of $\sigma \in \cA$, denoted $\Ostar{\sigma}$, is the union of the interiors
of all simplices in $\cA$ containing $\sigma$.
See Fig.~\ref{Fig-StandardConstructions}.

A subdivision of a complex $\cA$ is a complex $\cB$ such that:
(i) for \emph{any} $\tau \in \cB$, $|\tau|$ is contained in
the polyhedron of some $\sigma \in \cA$.
(ii) for \emph{any} $\sigma \in \cA$, $|\sigma|$ is
the union of disjoint polyhedrons of simplices belonging to $\cB$.
We understand the subdivision as an operator $\Div$ from complexes to complexes.
If we perform $N$ consecutive applications of $\Div$, the composite operator is denoted by $\Div^N$.

A \emph{mesh-shrinking} subdivision $\Div \cA$ of a complex $\cA$
is a subdivision where, for any $1$-simplex $\sigma \in \skel^1(\cA)$,
$\Div \sigma$ contains at least two distinct $1$-simplices\footnote{
%ALTERNATIVE DEFINITION: such that the longest polyhedron of an edge in $\Div\cA$
%                        (i.e., a $1$-simplex in $\skel^1(\Div\cA)$)
%                        is strictly smaller than the corresponding one in $\cA$ \footnote{
Intuitively, we are ``shrinking'' the simplices.}.
A particularly important mesh-shrinking subdivision in this work
is the \emph{barycentric subdivision}.

\begin{figure}[!htb]
\centering{\includegraphics[width=0.6\textwidth]{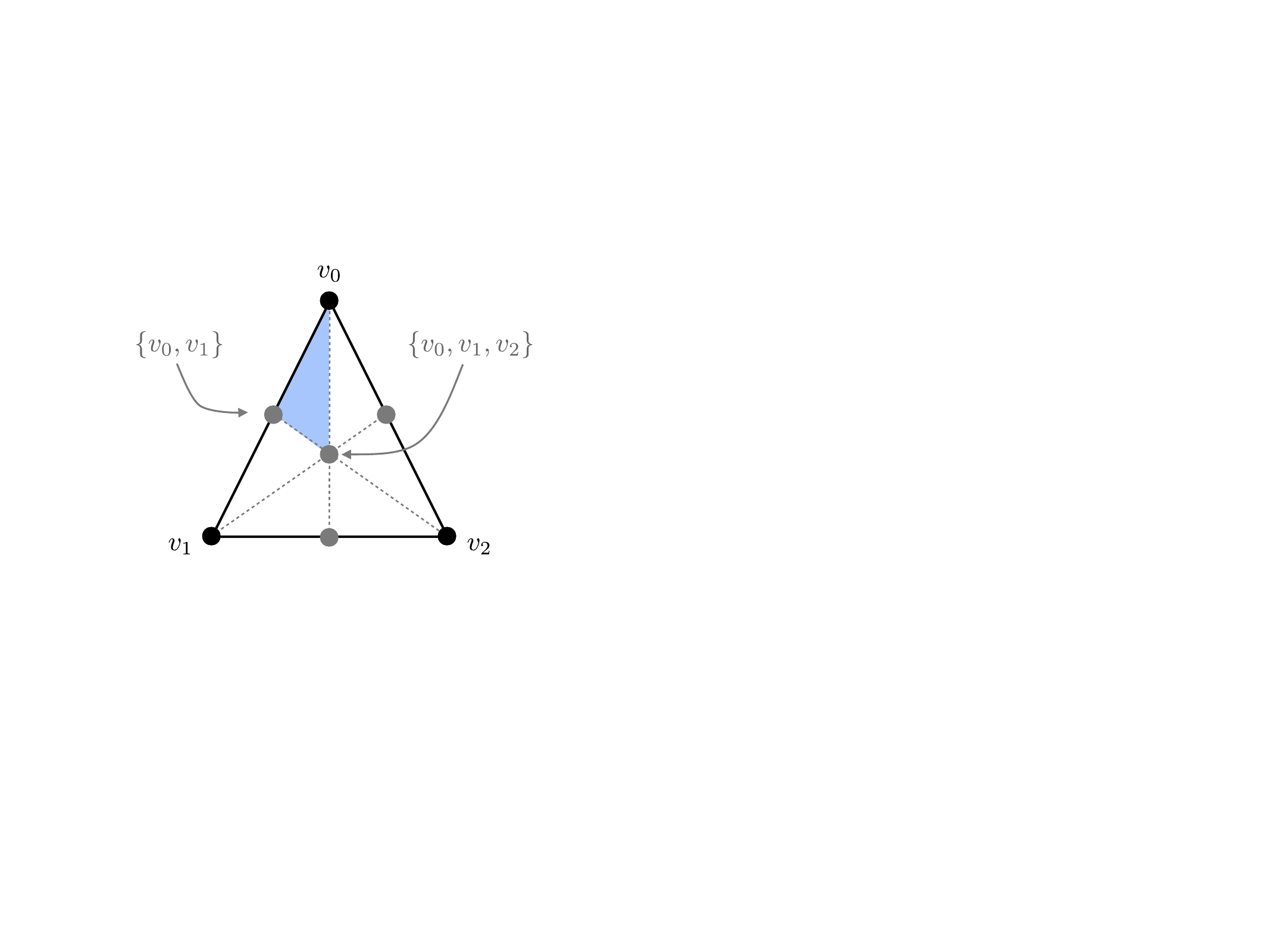}}
\caption{Barycentric subdivision of $\sigma = \set{v_0, v_1, v_2}$,
and one of its simplices $\set{\set{v_0}, \set{v_0,v_1}, \set{v_0,v_1,v_2}}$.}
\label{Fig-BarycentricSubdivision}
\end{figure}

Formally,
the barycentric subdivision of $\sigma$ is a simplicial complex $\bary{\sigma}$
whose vertices are faces of $\sigma$,
and whose simplices are chains of distinct faces totally ordered by containment.
Every $m$-simplex $\tau \in \bary{\sigma}$ might be written as
$\set{\sigma_0, \ldots, \sigma_m}$,
where $\sigma_0 \subset \cdots \subset \sigma_m \subseteq \sigma$.
See Fig.~\ref{Fig-BarycentricSubdivision}.

Given any continuous map, a simplicial approximation is,
intuitively speaking,
a ``sufficiently close'' combinatorial counterpart.
In accordance with \cite{Munkres84,Kozlov07},
we give a formal definition:
\begin{definition}
\label{definition:simpapprox}
A simplicial map $\mu: \cK \to \cL$ is a \emph{simplicial approximation}
of $\Phi_c: |\cK| \to |\cL|$ if
\begin{displaymath}
\Phi_c(|\inte{\sigma}|) \subseteq \bigcap_{v \in \sigma} |\Ostar{\mu(v)}| = |\Ostar{\mu(\sigma)}|, \textrm{ for all $\sigma \in \cK$.}
\end{displaymath}
\end{definition}

\begin{theorem}
\label{theorem:simpapprox}{(Simplicial Approximation, \cite{Munkres84,Kozlov07})}
For any continuous map $\Phi_c: |\cK| \to |\cL|$, consider an arbitrary
mesh-shrinking subdivision operator $\Div$.
Then, there exists an $N > 0$ such that $\Phi_c$ has a simplicial approximation $$\mu: \Div^N \cK \to \cL.$$
\end{theorem}

%For details, please refer to~\cite{Munkres84,Kozlov07}.

\subsection{Connectivity}
\label{App-Connectivity}

We say that a simplicial complex $\cK$
is $x$-connected
if every continuous map
of a $x$-sphere in $|\cK|$ can be extended into
the continuous map of a $(x+1)$-disk in $|\cK|$.
In analogy, think of a pencil as a $1$-disk,
and its extremes as a $0$-sphere;
a coin as a $2$-disk,
and its border as a $1$-sphere;
a billiard ball as a $3$-disk,
and its outer layer as a $2$-sphere.
In addition, $(-1)$-connected is defined as non-empty.

\begin{fact}
\label{fact:simplexhomomorphism}{\cite{Munkres84,Kozlov07}}
For any $k$-simplex $\sigma$,
the boundary of $\sigma$ is homeomorphic to a $(k-1)$-sphere,
and $\sigma$ is homeomorphic to a $k$-disk.
\end{fact}

\subsection{Protocols and Complexes}
\label{App-ProtocolsComplexes}

We abstract protocols as continuous exchanges of internal states,
in a full-information fashion (\cite{HerlihyRT09} and Sec.~\ref{Sec-OperationalModel}).
We will also use simplicial complexes to model protocol executions. 
Given a model for communication and failures,
we can define a \emph{protocol complex} $\cP(\cI)$ for any task $(\cI,\cO,\Delta)$.
A vertex in $v \in \cP(\cI)$ is a tuple with a non-faulty process identifier and its final state.
A simplex $\sigma = \set{(Q_1,s_1), \ldots, (Q_x,s_x)}$ in $\cP(\cI)$ indicates that,
in some execution,
non-faulty processes $Q_1, \ldots, Q_x$ finish with states $s_1, \ldots, s_x$, respectively.
The formal definition of \emph{protocol} is identical to \cite{HerlihyR12},
here presented for completeness:

\begin{definition}
\label{definition:protocol}
A protocol for $(\cI,\cO,\Delta)$ is a carrier map $\cP$ that takes $\sigma \in \cI$
to a protocol complex that we denote by $\cP(\sigma) \subseteq \cP(\cI)$.
%For any $\sigma,\tau \in \cI$, 
%we have that $\cP(\sigma \cap \tau) = \cP(\sigma) \cap \cP(\tau)$.
\end{definition}

\begin{definition}
\label{definition:protosolve}
A protocol $\cP$ solves $(\cI,\cO,\Delta)$
if there exists a simplicial map $\delta: \cP(\cI) \to \cO$
carried by $\Delta$.
\end{definition}

\subsection{Solvability for Colorless Tasks}
\label{Sec-SolvabilityColorlessTasks}

In this section,
we explore our colorless model and the concepts above
to obtain computability conditions
specific to colorless tasks.
We start with an interesting consequence of having an asynchronous Byzantine protocol.

\begin{theorem}
\label{theorem-protomap}
If a colorless $(\csI, \csO, \sDelta)$ has a $t$-resilient protocol
in asynchronous Byzantine systems,
there exists a continuous map $f: |\skel^t(\csI)| \to |\csO|$ carried by $\sDelta$.
\end{theorem}
\begin{proof}
Assuming a protocol,
we argue by reduction to the crash-failure case,
and then proceeding similarly to~\cite{HerlihyR10}.
First, note that any $t$-resilient Byzantine protocol is also
a $t$-resilient crash-failure protocol.
From~\cite{HerlihyR10Shellable,HerlihyRT09}\footnote{
These papers characterize connectivity in terms of the minimum core size $c$,
as defined by Junqueira and Marzullo \cite{JunqueiraM03}.
For $t$-resilient tasks
in the crash-failure model, $t = c+1$.
%See also Appendix~\ref{App-SolvabilityNonIndependentFaults}.
},
for any $\sigma \in \csI$, the protocol complex $\cP(\sigma)$
is $(t-1)$-connected in the crash-failure model,
so, in light of the previous observation,
it is also $(t-1)$-connected in the Byzantine-failure model,
with processes failing by crashing.
This implies that $\skel^x(\cP(\sigma))$ is $(x-1)$-connected
for $0 \le x \le t$.
We will then inductively construct a sequence of continuous maps
$g_x: |\skel^x(\csI)| \to |\cP(\skel^x(\csI))|$, for $0 \le x \le t$,
mapping skeletons of $\csI$ to skeletons of $\cP(\csI)$ as in~\cite{HerlihyR10}.

\textbf{Base.} Let $g_0$ map any vertex $v \in \sigma$ to any vertex $v' \in \cP(v)$,
which exists because $\skel^0(\cP(v))$ is $(-1)$-connected by hypothesis.
We just constructed
$$ g_0: |\skel^0(\csI)| \to |\cP(\skel^0(\csI))| \mathrm{.} $$

\textbf{Induction Hypothesis.} Assume
$$ g_{x-1}: |\skel^{x-1}(\csI)| \to |\cP(\skel^{x-1}(\csI))| \mathrm{,} $$
with $x \le t$,
sending the geometrical boundary of a $x$-simplex $\sigma^x$
in $\skel^x(\csI)$ to $|\cP(\skel^{x-1}(\sigma^x))|$.
In other words, we have $g_{x-1}(|\Bd\sigma^x|) \subseteq |\cP(\skel^{x-1}(\sigma^x))|$.
By hypothesis, $\cP(\sigma^x)$ is $x$-connected,
so the continuous image of the $(x-1)$-sphere $|\Bd\sigma^x|$ could be extended to
a continuous $x$-disk $|\sigma^x|$,
defining $g_x$ such that 
$g_x(|\sigma^x|) \subseteq |\cP(\skel^x(\csI))|$.
As all such maps agree on their intersections,
we just constructed
$$ g_x: |\skel^x(\csI)| \to |\cP(\skel^x(\csI))| \subseteq |\cP(\csI)| \mathrm{.} $$

As we assumed a protocol solving $(\csI,\csO,\sDelta)$,
we have a simplicial map $\delta^*: \cP(\csI) \to \csO$ carried by~$\sDelta$
(given by Definition~\ref{definition:protosolve}).
Our map is induced by the composition $\delta^*\circ g_t $.
For details, on the induced composition, please refer to~\cite{Munkres84,Kozlov07}.
\end{proof}

Colorless tasks have varying requirements
in terms of the number of processes required for solvability.
Consider strict $(t+1)$-set agreement.
If $\dim(\csI) \le t$ (which includes the case when $\dim(\csI) = 0$),
each process can simply decide on its input, without any communication.
For non-trivial cases,
the protocol requires $n + 1 > t(\dim(\csI) + 2)$,
shown in Lemma~\ref{lemma-procskset}.
The result follows
as an application of our Equivalence Theorem (Theorem~\ref{theorem-equivalence}).

\begin{lemma}
\label{lemma-procskset}
The strict $(t+1)$-set agreement task $$\cT = (\csI,\csO,\skel^t) \mathrm{,}$$
has a $t$-resilient protocol in asynchronous Byzantine systems
if and only if $n + 1 > t(\dim(\csI) + 2)$ or $\dim(\csI) \le t$.
\end{lemma}
\begin{proof}

($\Leftarrow$)
If $\dim(\csI) \le t$, a $k$-set agreement protocol is trivial,
as non-faulty processes already start with at most $t+1$ distinct values in~$\csI$.
Otherwise,
in the situation where $(n+1)-t > t(\dim(\csI) + 1)$,
consider Alg.~\ref{Alg-KSetStrictAgree}.
Assuming for contradiction that each of the $\dim(\csI)+1$ input values
is chosen by at most $t$ different non-faulty processes,
we would have that $(n+1)-t \le t(\dim(\csI)+1)$.
Therefore, at least $t+1$ non-faulty processes in fact input an identical value $v$,
and non-faulty processes can wait for such occurrence,
eventually deciding on a value inside $\csI$.
Also, as $(n+1)-t$ messages are received via reliable broadcast,
at most $t$ values are missed,
so at most $t+1$ values are decided,
solving the problem.
\begin{algorithm}[H]
\caption{$P_i.\mathrm{KSetStrictAgree}(I_i)$}
\label{Alg-KSetStrictAgree}
\begin{algorithmic}[1]
\If{$\dim(\csI) \le t$}
	\Return $I_i$
\EndIf
\State Get $(n+1)-t$ messages with values in $\csI$ via reliable broadcast, with some value appearing $t+1$ times
\State \Return $O_i = $ the smallest value received
\end{algorithmic}
\end{algorithm}

($\Rightarrow$)
If $\cT$ is solvable,
then $\cT' = (\sigma^*, \sigma^*, \skel^t)$ is similarly solvable,
taking an arbitrary $d$-simplex $\sigma^* = \set{v_0, \ldots, v_d}$ in $\csI$
with $d = \dim(\csI)$.
By our equivalence theorem,
and considering the relations in~(\ref{eqn-eqvcoloredcolorless1}) and~(\ref{eqn-eqvcoloredcolorless2}),
we must be able to solve
\begin{align}
\cT'' = (\Psi(\sigma^*;\bbP),\Psi(\sigma^*;\bbP),\widetilde{\skel^t}) \mathrm{,}
\end{align}
where, for any canonical name-labeled $\sigma,\tau \in \Psi(\sigma^*;\bbP)$:
\begin{multline}
\tau \in \widetilde{\skel^t}(\sigma) \Leftrightarrow
\forall \textrm{ canonical } \sigma' \subseteq \sigma,
\exists \textrm{ matching } \tau' \subseteq \tau \\
\textrm{ with }
\views(\tau') \in \skel^t(\views(\sigma')) \mathrm{.}
\end{multline}

Assume a protocol for $\cT''$,
and for contradiction, assume that $(n + 1) - t \le t(\dim(\csI) + 1)$.
Consider an execution where:
(i) all processes behave correctly or crash;
(ii) all processes in $\bbS = \set{P_0, \ldots, P_{n-t}}$ terminate without
receiving any message from any process in $\bbT = \set{P_{n+1-t}, \ldots, P_n}$;
(iii) each process $P_i \in \bbS$ starts with input $I_i = v_{i \bmod d + 1}$.
In this case, define
\begin{equation}
\label{equation:defsets}
\bbS_x = \set{p \in \bbS : p \text{ has input } v_x}.
\end{equation}
Note that $(n + 1) - t \ge d + 1$,
as $\csI$ contains only inputs chosen by non-faulty processes
and $d = \dim(\csI)$ by assumption.
Consequently, since $(d + 1) \le n + 1 - t \le t (d + 1)$,
and in light of (\ref{equation:defsets}),
\begin{equation*}
\label{equation:sizesets}
0 < |\bbS_x| \le t \textrm{ for all } 0 \le x \le d \mathrm{.}
\end{equation*}
Regarding notation,
we define $\sigma_{x} = \set{(P_i,I_i): P_i \in \bbS_x}$ and
$\sigma_{-x} = \set{(P_i,I_i): P_i \in \bbS \setminus \bbS_x}$, concerning the inputs;
also $\tau_{x} = \set{(P_i,O_i): P_i \in \bbS_x}$ and
$\tau_{-x} = \set{(P_i,O_i): P_i \in \bbS \setminus \bbS_x}$, concerning the outputs.

In order to satisfy $\widetilde{\skel^t}$,
given that all processes in $\bbS$ decide,
and that the values of the processes in $\bbT$ are unknown,
we must have:
$$\views(\tau_{-x}) \in \skel^t(\views(\sigma_{-x})) = \skel^t(\sigma^* - \set{v_x}) \mathrm{.}$$
Therefore,
\begin{align*}
\views(\tau_{x}) & \subseteq \bigcap_{y \ne x} \skel^t(\views(\sigma_{-y})) \\
                 & = \bigcap_{y \ne x} \skel^t(\sigma^* - \set{v_y}) \\
								 & \subseteq \set{v_x} \mathrm{,}
\end{align*}
for any $0 \le x \le n-t$.
In conclusion,
each process decides its own input,
and the decision fails to solve the problem unless
$d + 1 \le t + 1$,
which implies $\dim(\cI) \le t$.
In the latter situation,
the protocol is trivial: each process can choose its own input without any communication.
\end{proof}

The previous requirement on the number of processes,
although not a necessary condition for the solvability of all colorless tasks,
is part of an interesting sufficient condition for solvability.
We shall consider the interesting cases where  $\dim(\csI) > 0$.

\begin{theorem}
\label{theorem-mapproto}
For any colorless $\cT = (\csI,\csO,\sDelta)$, if
\begin{enumerate}
	\item $n + 1 > t(\dim(\csI) + 2)$; and
	\item there exists a continuous map $f: |\skel^t(\csI)| \to |\csO|$ carried by $\sDelta$,
\end{enumerate}
then we have a $t$-resilient protocol in asynchronous Byzantine systems for $\cT$.
\end{theorem}
\begin{proof}
By the simplicial approximation theorem \cite{Munkres84,Kozlov07}
(see Theorem~\ref{theorem:simpapprox}),
$f$ has a \emph{simplicial approximation}
$$\phi: \bary^N{\skel^t(\csI)} \rightarrow \csO \mathrm{,}$$ for some $N > 0$,
also carried by $\sDelta$.
The Byzantine-failure protocol for non-faulty processes is shown below,
presuming that $n + 1 > t(\dim(\csI) + 2)$ with $\dim(\csI) > 0$.
\begin{enumerate}
\item
Execute the Byzantine strict $(t+1)$-set agreement protocol (Algorithm~\ref{Alg-KSetStrictAgree}),
choosing vertices on a simplex in $\skel^t(\csI)$.
\item
Execute $N$ times the Byzantine barycentric agreement protocol,
choosing vertices on a simplex in $$\bary^N{\skel^t(\csI)} \mathrm{.}$$
For simplicity of presentation,
we assume the approach described in Appendix~\ref{App-BaryAgreeApproxAgree},
based on~\cite{MendesHerlihy13}.
\item
Apply $\phi: \bary^N{\skel^t(\csI)} \to \csO$ to choose vertices on a simplex in $\csO$.
\end{enumerate}
As $\phi$ and $f$ are carried by $\sDelta$,
non-faulty processes starting on vertices of $\sigma_I \in \csI$ finish on
vertices of $\sigma_O \in \sDelta(\sigma)$.
Furthermore, since $1 \le \dim(\sigma_I) \le \dim(\csI)$,
by definition,
the preconditions are satisfied for calling the protocols
in steps (1) and (2).
\end{proof}

\section{Conclusion}
\label{Sec-Conclusion}

In this work,
we give novel necessary and sufficient conditions for task solvability
in asynchronous Byzantine systems.
While analogous results have long existed for crash-failure systems~\cite{HerlihyShavit1999},
we provide solvability conditions for arbitrary Byzantine tasks for the first time.
Here, we presume an adversarial model to specify which processes are Byzantine.
Independently of which processes are deemed faulty,
any final configuration of the non-faulty processes
must be permitted in respect to
the initial configuration of the non-faulty processes,
according to a formal task specification.

For Byzantine colorless tasks,
a specialized, more fitting model permits us to express slightly more specific conditions for solvability.
In particular,
we show that the strict $k$-set agreement requires a certain number of processes,
except in trivial cases.
Furthermore,
we show that a colorless Byzantine protocol under asynchronous systems
implies the existence of a continuous map $f: |\skel^t \csI| \to |\csO|$ carried by
the carrier map $\sDelta$,
and that $n + 1 > t(\dim(\csI) + 2)$ is enough to solve an arbitrary colorless task when such map indeed exists.

In this work,
we demonstrate how the language and techniques of combinatorial topology
can produce novel results in distributed computing,
facilitating existential arguments
while avoiding complicated, model-specific argumentation.
As future work,
we intend to explore other requirements for solvability in colorless tasks,
specifically in regard to the number of processes,
perhaps unifying Theorems~\ref{theorem-protomap} and~\ref{theorem-mapproto}.

We thank Petr Kuznetsov, Zohir Bouzid, and Eli Gafni for comments on previous versions of this work.

\bibliography{Bibliography.bib}
\bibliographystyle{abbrv}

\appendix

%On this Appendix,
%we present algorithms for the reliable broadcast protocol
%and an algorithm to solve the barycentric agreement problem
%via the multidimensional $\epsilon$-approximate agreement problem of~\cite{MendesHerlihy13}.

\section{Reliable Broadcast Protocol}
\label{App-ReliableBroadcast}

Algorithms~\ref{Alg-RB1}~to~\ref{Alg-RB3} show the
reliable broadcast protocol for
sender $P$, round $r$, and content $c$.
The symbol ``$\cdot$'' represents a wildcard,
matching any value.
For proofs and details,
please refer to~\cite{DSBook,RSPBook,Bracha}.

\begin{algorithm}[H]
\caption{$P.\mathtt{RBSend}((P,r,c))$}
\label{Alg-RB1}
\begin{algorithmic}
  \State \send $(P,r,c)$ to all processes
\end{algorithmic}
\end{algorithm}

\begin{algorithm}[H]
\caption{$P.\mathtt{RBEcho}()$}
\label{Alg-RB2}
\begin{algorithmic}
  \Upon{\recv $(Q,r,c)$ from $Q$}
    \If{never sent $(P,Qr \mathtt{\{echo\}},\cdot)$}
      \State \send $(P,Qr \mathtt{\{echo\}},c)$ to all processes
    \EndIf
  \EndUpon

  \Upon{\recv $(\cdot,Qr \mathtt{\{echo\}},c)$ from $\ge n+1-t$ processes}
    \If{never sent $(P,Qr \mathtt{\{ready\}},\cdot)$}
      \State \send $(P,Qr \mathtt{\{ready\}},c)$ to all processes
    \EndIf
  \EndUpon

  \Upon{\recv $(\cdot,Qr \mathtt{\{ready\}},c)$ from $\ge t + 1$ processes}
    \If{never sent $(P,Qr \mathtt{\{ready\}},\cdot)$}
      \State \send $(P,Qr \mathtt{\{ready\}},c)$ to all processes
    \EndIf
  \EndUpon
\end{algorithmic}
\end{algorithm}

\begin{algorithm}[H]
\caption{$P.\mathtt{RBRecv}((Q,r,c))$}
\label{Alg-RB3}
\begin{algorithmic}
  \State \recv $(\cdot,Qr \mathtt{\{ready\}},c)$ from $(n+1)-t$ processes
  \State \Return $(Q,r,c)$
\end{algorithmic}
\end{algorithm}

\section{Barycentric Agreement via Approximate Agreement}
\label{App-BaryAgreeApproxAgree}

In this section, we show how to transform a protocol
for the multidimensional $\epsilon$-approximate agreement problem,
defined in~\cite{MendesHerlihy13},
into barycentric agreement.

The multidimensional $\epsilon$-approximate agreement task is defined as follows.
Consider a set of $n+1$ processes,
including no more than $t$ Byzantine processes.
Every process $P_i \in \bbG$ has an input $I_i \in \bbR^m$ and an output $O_i \in \bbR^m$.
After we run the protocol, we require:
\begin{description}
	\item[Agreement:] for any non-faulty processes $P_i$ and $P_j$, the Euclidean distance between their outputs $O_i$ and $O_j$ is $\le \epsilon$, an error tolerance fixed \emph{a priori}.
	\item[Convexity:] for any non-faulty process $P_i$, its output $O_i$ is in the convex hull of the inputs of the non-faulty processes.
\end{description}

The point-set occupied by $\cI$ is compact, and the open stars of the
vertices of $\bary \cI$ form an open cover of $\cI$.
Any such cover has a \emph{Lebesgue} number $\lambda > 0$~\cite{Munkres84},
such that every set of diameter less than $\lambda$ is contained in
some member of the cover.

We now describe a Byzantine barycentric agreement protocol.
Suppose the non-faulty processes start at the vertices of an input simplex
$\sigma$.
Using $(\lambda/2)$-approximate agreement,
each non-faulty process $p_i$ chooses a point inside $\sigma$,
the convex hull of the inputs,
such that the distance between any pair of points is less than $\lambda/2$.
Equivalently,
each open ball of radius $\lambda/2$ around $v_i$
contains all values chosen by the approximate agreement protocol.
Because the diameter of this set is less than the Lebesgue number $\lambda$,
there is at least one vertex $u_i$ in $\bary \cI$ such that
$B(v_i,\lambda/2)$ lies in the open star around $u_i$.
Let each $P_i$ choose any such $u_i$.

We must still show that the vertices $u_i$ that were chosen by the processes $P_i$ lie on a single
simplex of $\bary \sigma$.
Note that $u_i,u_j$ are vertices of a common simplex if and only if
the open star around $u_i$ intersects the open star around $u_j$.
By construction,
$v_j \in B(v_i,\lambda/2)$,
which is in the open star around $u_i$,
and $v_j$ is in the open star around $u_j$,
hence $u_i,u_j$ are vertices of a single simplex.

\end{document}